\documentclass[letterpaper, 10 pt, conference]{ieeeconf}  
                     
\usepackage{fancyhdr}
\usepackage{graphicx}
\usepackage{cite}
\usepackage{hyperref}
\usepackage{subfig,float}
\usepackage{epsfig}

\IEEEoverridecommandlockouts                              
\usepackage[utf8]{inputenc}
\usepackage[T1]{fontenc}

\title{\LARGE \bf
Interpretable SincNet-based Deep Learning  for \\ Emotion Recognition from EEG brain activity
}

\author{Juan Manuel Mayor-Torres$^{1}$, Mirco Ravanelli$^{2}$, Sara E. Medina-DeVilliers$^{3}$, Matthew D. Lerner$^{3}$, \\ and Giuseppe Riccardi$^{1}$
\thanks{}
\thanks{$^{1}$University of Trento, Department of Information Engineering and Computer Science (DISI), Via Sommarive 5, Povo, Trento 38123,
$^{2}$Mila - Quebec Artifical Intelligence Institute, Montreal, QC, Canada,
       $^{3}$ StonyBrook University, NY, USA, Department of Psychology 
        {\tt\small corresponding author: juan.mayortorres@unitn.it}}%
}

\begin{document}

\maketitle
\thispagestyle{empty}
\pagestyle{empty}

\begin{abstract}
Machine learning methods, such as deep learning, show  promising  results  in  the  medical  domain.  However,  the lack of interpretability of these algorithms may hinder  their applicability to medical  decision support systems.  This  paper  studies  an  interpretable deep  learning  technique,  called  SincNet.  SincNet  is  a  convolutional  neural  network  that  efficiently  learns  customized  band-pass  filters  through  trainable  sinc-functions.  In  this  study,  we use  SincNet  to  analyze  the  neural  activity  of  individuals  with Autism  Spectrum  Disorder  (ASD), who experience characteristic differences in neural oscillatory activity.  In  particular,  we  propose a  novel  SincNet-based  neural  network  for  detecting  emotions in ASD patients using EEG signals. The learned filters can be easily inspected  to  detect  which  part  of  the  EEG  spectrum is  used  for  predicting  emotions.  We  found  that  our  system automatically  learns  the  high-$\alpha$ (9-13  Hz)  and $\beta$ (13-30  Hz) band  suppression  often  present  in  individuals  with  ASD. This result is consistent with recent neuroscience studies on emotion recognition,  which  found  an  association  between  these band suppressions and the behavioral deficits observed in individuals with ASD. The improved interpretability of SincNet is achieved without  sacrificing  performance  in  emotion  recognition.  
\end{abstract}
\begin{figure*}[t!]
\centering
\includegraphics[width=16.5cm,height=5.2cm]{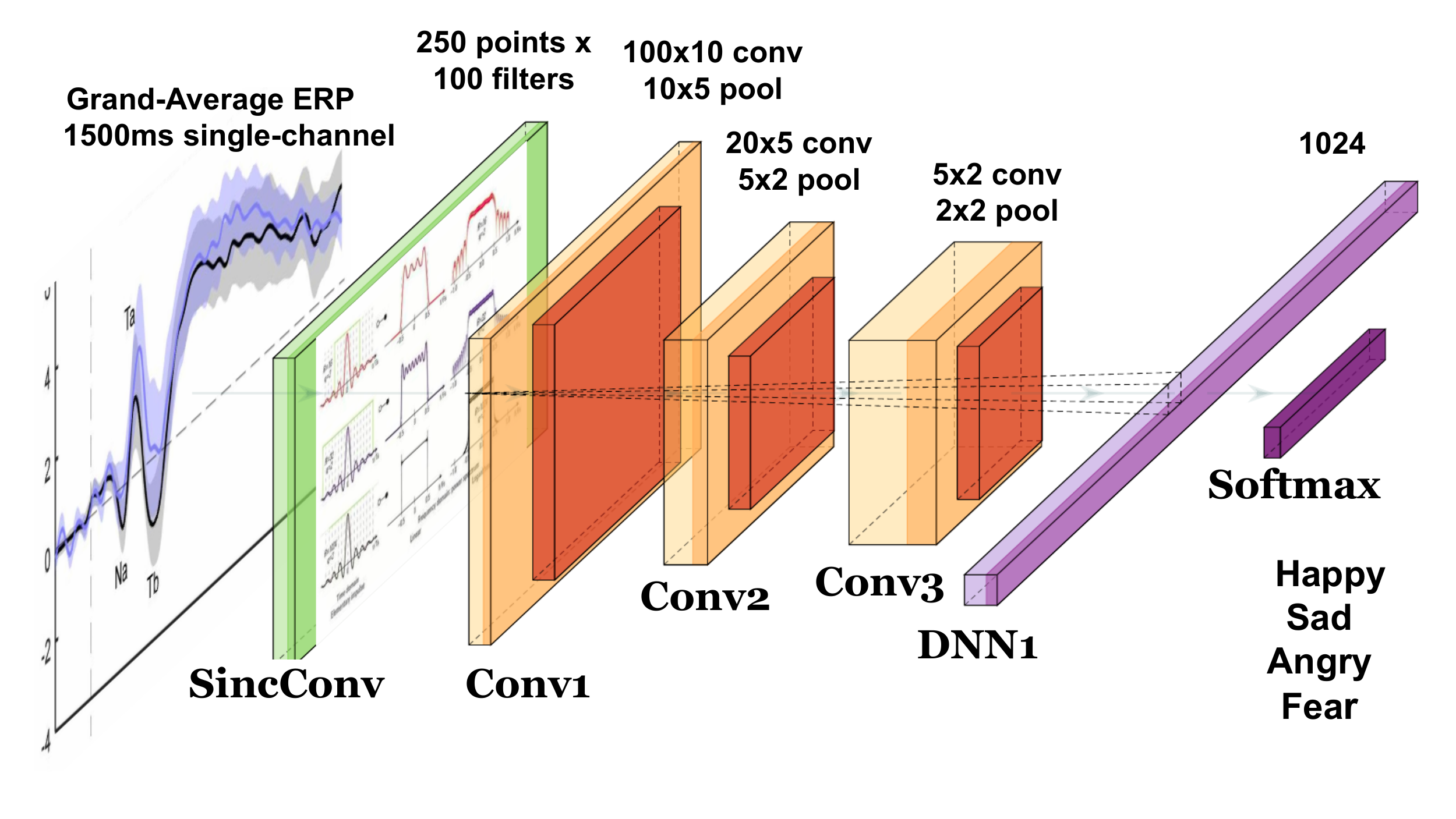}
\caption{\small{The proposed architecture for EEG-based Face Emotion Recognition. The neural pipeline is composed of a \emph{SincConv} layer, three standard conv-pool blocks (\emph{Conv1}, \emph{Conv2}, and \emph{Conv3}), and a fully-connected network (\emph{DNN1}) connected to a softmax classifier.}}
\label{fig:pipe}
\end{figure*}
\section{INTRODUCTION}
The recent development of artificial intelligence fosters unprecedented innovation in the healthcare domain. The availability of big data repositories, the development of robust learning algorithms, and the availability of appropriate computational resources are making technologies such as deep learning, applicable and challenging the state-of-the-art systems.
Deep learning (DL) has indeed been used in many medical applications \cite{torres2018enhanced}, including disease diagnosis \cite{Oh2018ADL}, and personalized medicine \cite{Papadakis}, just to name a few.
In such critical cases, the lack of interpretability of this technology limits its widespread  end-user adoption and may even lead to adverse consequences \cite{Aurangzeb}.
\newline
Current deep neural networks map low-level data into higher-level concepts using a pipeline of non-linear transformations \cite{goodfellow2016deep}. Therefore, the final prediction is normally not explained by the network, and  end-users ( i.e. healthcare professionals ) do not know if the outcome is based on solid evidence or due to some statistical biases \cite{lapuschkin2019unmasking}. Moreover, the inspection of the intermediate representations learned by the network rarely helps to explain the neural predictions. Improving the interpretability of the current technology is an essential condition for overcoming understandable skepticism, reluctance, and hesitations of the medical personnel.
\newline
Interpretable deep learning has been the object of increasing research efforts over the last years \cite{molnar2019}.
Historically, there was a trade-off between performance and interpretability. Simple models like linear regression are transparent but not competitive in terms of performance with deeper models such as fully-connected, convolutional, and recurrent neural networks.  A  possible  solution may be  post-hoc  interpretability \cite{molnar2019}, in which, a complex  neural  model is built and  analyzed  afterward .  The interpretability can also be achieved with surrogate models (e.g., locally interpretable model explanations - LIME \cite{ribeiro_lime}) and with gradient-based methods like in the saliency maps \cite{Simonyan}. In contrast, the alternative is to train machine learning models that are  interpretable by design. Along this line,  a novel model called SincNet has been proposed \cite{ravanelli2018speaker}. SincNet is a convolutional neural network that learns a custom filter-bank using sinc-based convolutional kernels.  Following the network training, the filters can be inspected in the frequency domain to identify which parts of the spectrum are used by the neural network to perform a prediction. The filter inspection is straightforward  and often insightful, as emerged in some recent studies \cite{ravanelli2018interpretable, zeng2019eeg,borra2020interpretable}. The interpretability insights are directly mapping the neural network parameters  to the input signal.
SincNet has been originally proposed for processing audio sequences \cite{ravanelli2018speaker}, but it has been recently used for EEG-based brain signals as well. In particular, SincNet has been successfully adopted for EEG-based motor-imagery tasks \cite{zeng2019eeg,borra2020interpretable}. The full potential of this model in high-level processing of input stimuli (e.g. visual or audio), however, is yet to be explored. 
\newline
In this paper, we propose SincNet for studying the brain activity of patients with Autism Spectrum Disorder (ASD). In particular, we analyze EEG signals of ASD and non-ASD individuals while performing a Facial Emotion Recognition (FER) task. 
The EEG recordings feed a deep learning model based on SincNet, which tries to guess patients' emotions from their brain waves. From the study of filters learned by SincNet, we found an interesting consistency  with  the  Power-Spectral-Density  analysis  of  previous EEG  studies on ASD. The improved interpretability is achieved without sacrificing the performance of the machine learning models. The proposed system provides a contribution to the interpretability-by-design alternative by applying deep learning techniques to the medical domain.
\section{Facial Emotion Recognition with SincNet}
The Facial Emotion Recognition (FER) task measures the ability to identify basic emotions in facial expressions. It is particular important to asses this ability in individuals with Autism Spectrum Disorder (ASD) because they frequently demonstrate impairments in the ability to accurately categorize and label the emotional facial expressions of others. In this section, we first describe the FER task. Then, we describe the proposed architecture for EEG-based emotion recognition in ASD and typically developing (TD) individuals.
\subsection{FER Task Description}
Eighty-eight 14-to-17-year old adolescent participants (48  without ASD, while  40  with  ASD) completed a FER task. 
The FER task, called Diagnostic  Analysis of Nonverbal Behavior (DANVA-2) \cite{nowicki2000manual}, consists of presenting 48 photographs in random order. Each photograph represents one of following emotions: \textit{happy}, \textit{sad}, \textit{angry}, and \textit{fear}. The picture was first shown to the users for 2 second and, within this interval, their EEG signal was recorded.  
Subsequently, each photograph was shown again (for up 4 seconds) to give the  participant  enough  time  to select the emotion they have seen in the picture using a button box. The EEG activity of  each  participant was fed into the proposed machine learning system, which is designed to predict the actual emotion from the given brain recordings. 
The full pipeline for EEG-based emotion recognition is shown in Figure \ref{fig:pipe}. 
\subsection{EEG pre-processing}
The EEG signals were captured with a  32-channel  ActiCHAMP device  from  Brain  Products  and  digitized  with a sampling rate of 500 Hz (16-bit  resolution).
The data were recorded continuously using the BrainVision Recorder software and processed using the BrainVision Analyzer 2.0 for offline data reduction. We converted all the input channels averaging them into a single sequence or a grand-average Event- Related Potential (ERP) representation for each participant and for each emotion. This helps reducing noise and statistical variability of EEG trials \cite{delorme2015grand}. 
We employed a standard EEG pre-processing pipeline based on bandpass filtering, amplitude normalization, and Zero Component Analysis (ZCA) whitening \cite{mayor2021biolpsych}. We also apply the ADJUST algorithm  \cite{mognon2011adjust} and the PREP pipeline \cite{bigdely2015prep} to clean the EEG signal before feeding the neural network.
\subsection{Intepretable SincNet}
The resulting grand-average ERP representation feeds the proposed SincNet-based system.
SincNet is a convolutional neural network whose first layer, called \textit{SincConv}, is designed to learn tunable Finite Impulse Response (FIR) filters.
In standard CNNs, all the elements (taps) of each filter are learned from the data. In SincConv, instead, we parametrize the kernels in order to implement rectangular bandpass filters, whose cut-off frequencies are the only two parameters learned from data. This can be achieved with the following parametrization:
\begin{equation}
y[n]=x[n]*g[n,f_1,f_2], 
\end{equation}
where $x[n]$ is the input EEG signal and $y[n]$ is the output of the SinConv layer. The convolution is performed with the kernel $g$, which is defined in this way: 
\begin{equation}
g[n,f_1,f_2]= 2f_{2}sinc(2\pi f_2n) - 2f_{1}sinc(2\pi f_1n),
\end{equation}
where the frequencies $f_1$ and $f_2$ are the learned parameters. This technique not only saves a lot of parameters but naturally leads to a more interpretable model. Moreover, the filters depend on human-readable parameters with a clear physical meaning. At the end of the training, the filters are inspected to identify which parts of the spectrum are covered by filters. This helps users better understand what the network has learned \cite{ravanelli2018interpretable, borra2020interpretable}. In addition to the SincConv layer, the proposed architecture employs three 2-D convolutional blocks based on standard convolution, batch normalization, pooling, ReLU activations, and dropout. Finally, we plug a fully connected layer followed by a softmax classifier.   
\section{EXPERIMENTAL SETUP}
In this section, we describe our SincNet-based emotion recognition architecture and its training procedure. 
\subsection{Architecture details}
The adopted SincConv layer employed 100 filters with a kernel size of 250 points (which represent the impulse response of the filter). As in the original paper, \cite{ravanelli2018speaker}, a Hamming window was used to mitigate ripples in the filters. The 2-D convolutional blocks were based on 32, 64, and 128 channels with kernel sizes of (100 x 10), (20 x 5), (5 x 2), respectively. Max-pooling used kernel sizes of (10 x 5), (5 x 2), (2 x 2). Batch normalization was added between the convolution and the ReLU activations. Next, we employed a single fully connected layer composed of 1024 ReLU neurons. Dropout was used in both convolutional and fully connected layers with a rate of 0.5.
The final prediction over the four emotions was performed with a softmax classifier. 
\subsection{Network Training}
We trained and evaluated the SincNet-based pipeline using a Leave-One-Trial-Out (LOTO) approach \cite{mayor2021biolpsych}. Thus, for each participant we used 47 out of the 48 trials for training, leaving a different test trial out every time. Therefore, we completed a total of 48 training/validation experiments. This modality was needed due to the lack of in-domain data for this specific task.
The neural network was initialized with the standard Glorot's initialization scheme \cite{glorot2010understanding}. We used categorical cross-entropy as a loss function. The gradient was computed with the backpropagation algorithm, while parameters are updated using the Adam optimizer \cite{kingma2014adam}. We used a learning rate of $0.001$ with a weight-decay penalty of $1e-05$. We trained the neural network for 400 epochs with a batch size of 30. For more information on the neural architecture and training modality, please refer to the open-source code repository of this project\footnote{\href{https://github.com/meiyor/SincNet-for-Autism-EEG-based-Emotion-Recognition}{https://github.com/meiyor/SincNet-for-Autism-EEG-based-Emotion-Recognition}}.
\begin{figure}[!t]
\hspace*{-0.4cm}
\includegraphics[width=9.7cm,height=4.6cm]{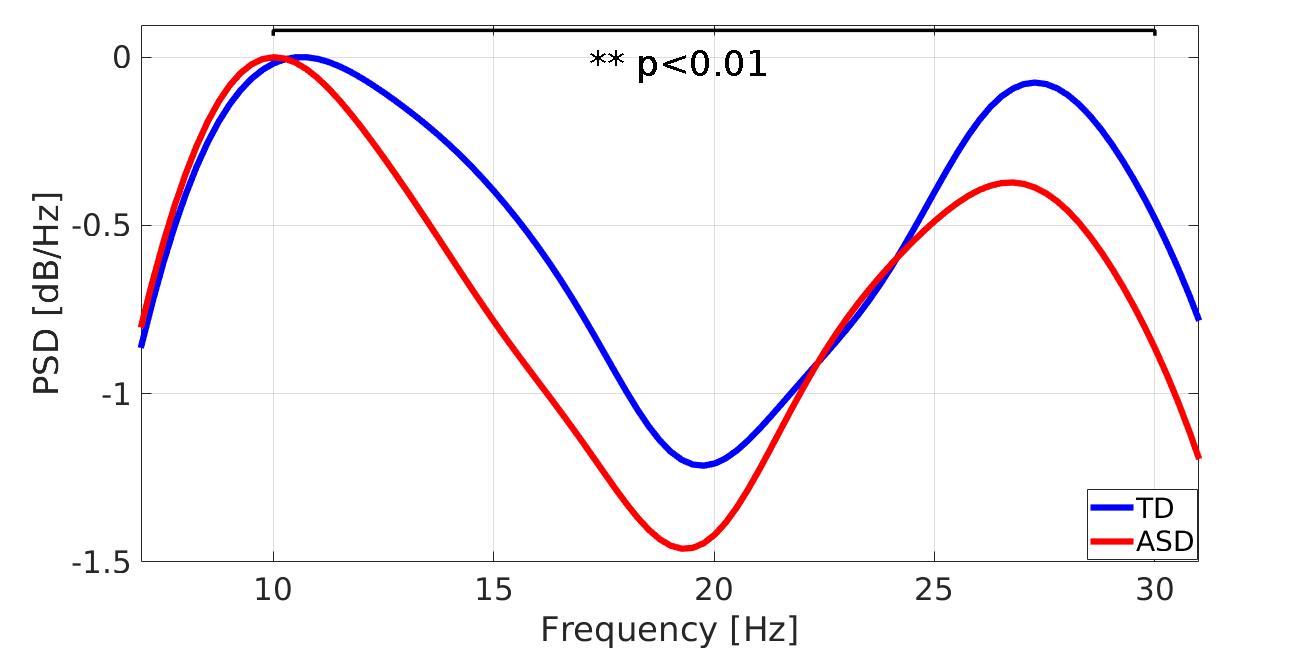}
\caption{\small{Power Spectral Density (PSD) learned by SincNet filters on TD and ASD individuals. Significant differences are observed in high-$\alpha$ (9-13 Hz) and $\beta$ (13-30 Hz) bands.}}
\label{fig:psd_learnt}
\end{figure}
\begin{figure}[t!]
\hspace*{-0.2cm}
\includegraphics[width=9.7cm,height=4.6cm]{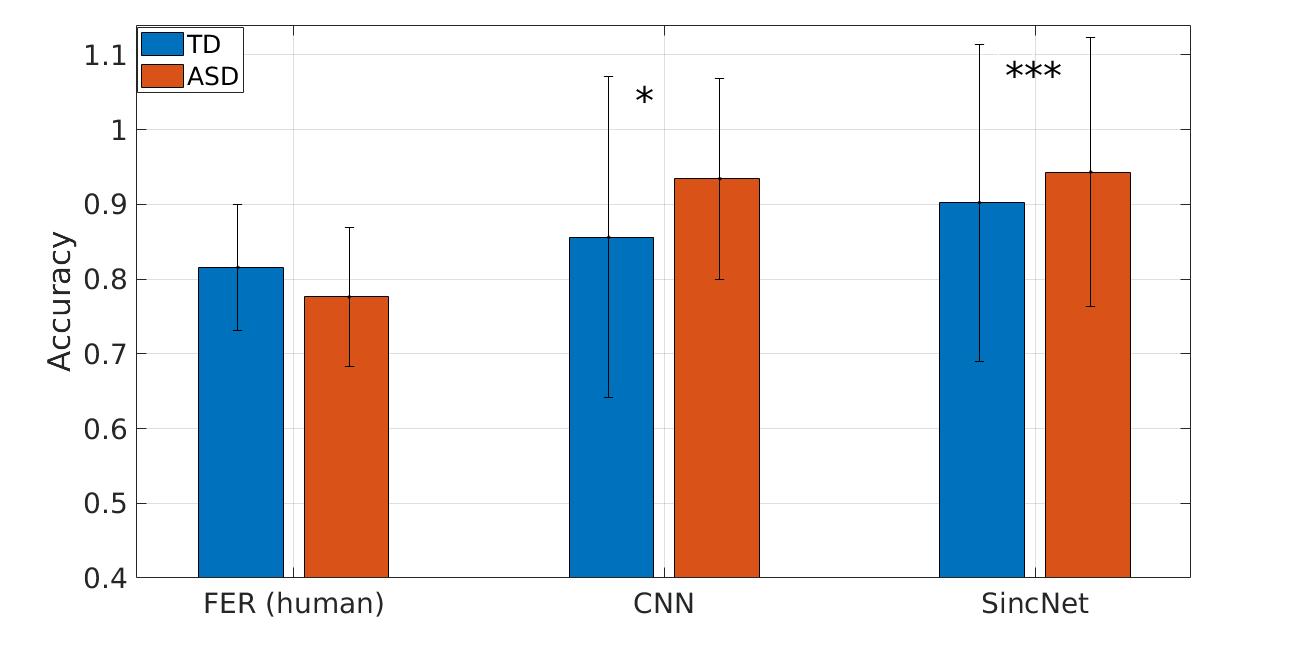}
\caption{\small{Accuracy comparison between FER (humans), CNN, and SincNet. Human accuracy is significantly lower than the one of deep learning systems.}}
\label{fig:performance}
\end{figure}
\section{RESULTS}
In the following section, we report the experimental evidence that emerged from the FER task using SincNet. 
\subsection{Filter Analysis}
After training,  we inspected the filters learned by SincNet for ASD and non-ASD (TD) participants. To analyze which EEG frequency bands were used for the emotion prediction, we performed a Fourier Transform of the learned filters. We then averaged their frequency responses and computed the cumulative Power Spectral Density (PSD) reported in Fig. \ref{fig:psd_learnt}.
We observed significant differences in the filters learned from ASD and non-ASD participants. In particular, an attenuation in the high-$\alpha$ (9-13 Hz) and $\beta$ (13-30 Hz) emerged in the cumulative PSD spectrum of ASD participants. We observed significant differences between non-ASD and ASD groups on high-$\alpha$ (9-13 Hz) $F(1,87)=3.331,p=0.000267$ and $\beta$ (13-30 Hz) $F(1,87)=2.05,p=0.00102$ bands after Bonferroni-Holm correction. Our findings are consistent with previous studies on ASD, indicating that high-$\alpha$ and $\beta$ attenuations are related to FER deficits on individuals with ASD \cite{pineda2012self,friedrich2015effective}.
\begin{figure}[t!]
\centering
\includegraphics[width=9.7cm,height=4.6cm]{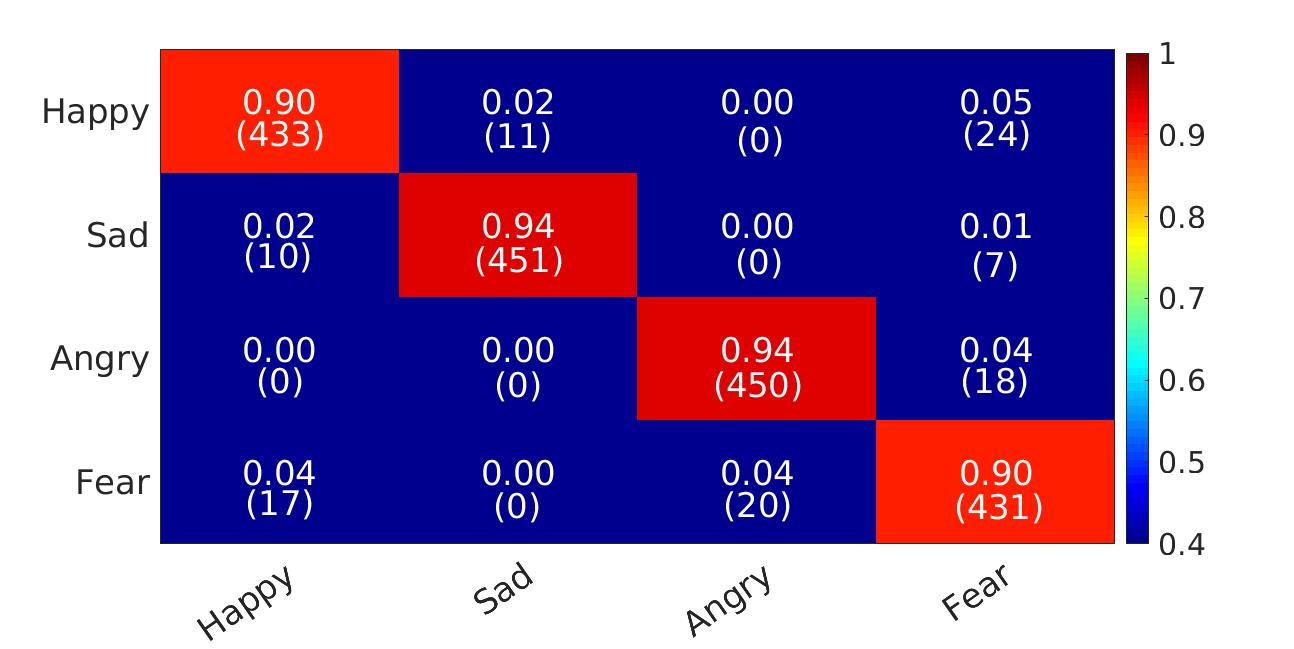}
\caption{\small{Confusion Matrix for the SincNet pipeline on ASD}.}
\label{conf_mat}
\end{figure}
Although this frequency-band attenuation has been observed in some EEG studies including individuals with ASD, the causes of this are not fully understood in the scientific community. Some authors hypothesized that this is associated with multiple behavioral deficits of individuals with ASD \cite{dawson2005understanding,black2017mechanisms}. However, it is worth notice that SincNet learns that these bands are not useful to predict emotions in individuals with ASD.  Notably, these predictions were learned automatically from raw EEG data only, without providing any additional information to the network.
\subsection{Performance Analysis}
Figure \ref{fig:performance} compares the accuracy achieved by humans in FER with the one reached by the deep learning systems based on CNNs and SincNet. 
As for the human performance or FER,  we found a small difference between ASD and non-ASD accuracies (79\% vs 81\%, respectively), thus, showing some consistent deficits in individuals with ASD performing FER tasks \cite{dawson2005understanding,black2017mechanisms}.
Interestingly, deep learning systems outperformed the FER human accuracy. This suggests that there were some cases where a participant wrongly labeled the photograph, but the SincNet-based system was able to detect the correct emotion from participants' EEG brain activity. 
This difference is more evident in ASD participants, where the SincNet improvement is greater (79\% vs 92\%) than FER. The SincNet confusion matrix for the ASD group is shown in Figure \ref{conf_mat}.
The proposed SincNet pipeline turned out to slightly outperforming a CNN-based system (90\% vs 85\% for the non-ASD group and 92\% vs 91\% for the ASD group). The CNN architecture was obtained by replacing the SincConv layer with a standard convolution.
This improvement is consistent with what was observed for audio \cite{ravanelli2018speaker} and motor-based EEG signal in previous studies \cite{borra2020interpretable}. In sum, our results confirm that SincNet improves the interpretability of the model without sacrificing performance. 

\section{CONCLUSIONS}
This paper has proposed the application of an interpretable deep-learning architecture, SincNet, in a medical domain.  
We applied this model to study EEG activity patterns of ASDs and non-ASDs in a FER task. Our results indicate that SincNet transparently learns the  high-$\alpha$ and $\beta$ suppressions observed in ASD individuals when perceiving and recognizing emotional faces. SincNet improves the interpretability of the neural model without affecting its performance, thus offering a convenient way to avoid the performance versus interpretability dilemma.
\section*{ACKNOWLEDGMENT}
The authors would like to thank the University of Trento High-Performance Computing and Stony Brook Research Computing and Cyber-infrastructure which was made possible by a \$1.4M National Science Foundation grant (1531492). This research was supported by NIMH grant R01MH110585, grants from the AAF Fund for Communication.
\bibliographystyle{unsrt}
\bibliography{reference}
\end{document}